\documentclass[10pt]{iopart}
\usepackage{graphicx}
\usepackage{color}

\begin{document}
\title[Evaluating the performance of the NPL femtosecond frequency combs]{Evaluating the performance of the NPL femtosecond frequency combs: Agreement at the $\mathbf{10^{-21}}$ level}
\author{L A M Johnson, P Gill and H S Margolis}
\address{National Physical Laboratory, Teddington, Middlesex, UK}
\ead{helen.margolis@npl.co.uk}

\begin{abstract}

\noindent Results are presented from a series of comparisons between two independent femtosecond frequency comb systems at NPL, which were carried out in order to assess their systematic uncertainty. 
Simultaneous measurements with the two systems demonstrate agreement at the level of 5$\times 10^{-18}$ when measuring an optical frequency against a common microwave reference. When simultaneously measuring the ratio of two optical frequencies, agreement at the 3$\times 10^{-21}$ level is observed.
The results represent the highest reported level of agreement to date between Ti:sapphire and Er-doped femtosecond combs. The limitations of the combs when operating in these two different manners are discussed, including traceability to the SI second, which can be achieved with an uncertainty below 1$\times 10^{-16}$.
The technical details presented underpin recent absolute frequency measurements of the $^{88}$Sr$^{+}$ and $^{171}$Yb$^{+}$ optical clock transitions at NPL, as well as a frequency ratio measurement between the two optical clock transitions in $^{171}$Yb$^{+}$.

\end{abstract} 
%\pacs{42.62.Eh, 06.30.Ft}

%\noindent{\it Keywords\/}: Frequency comb, optical frequency metrology, absolute frequency, optical frequency ratio

%\submitto{\MET}
\maketitle
\ioptwocol

\section{Introduction}

A wealth of recent research into next-generation optical atomic clocks based on trapped single ions and lattice-confined neutral atoms has resulted in such systems now outperforming the best Cs primary standards which currently realise the SI second. These optical clocks are opening up several interesting areas of research including improved tests of fundamental physics \cite{rosenband2008} and clock-based geodesy \cite{chou2010} and it is expected that a redefinition of the second in terms of an optical clock transition frequency will take place in the future. However, before such a redefinition can occur it is vital that different types of optical clock can be compared both against one another, to validate their uncertainty budgets, and against current Cs primary standards to relate them back to the present definition of the SI second. Femtosecond frequency combs are the tool used for both types of comparison.   

Although the fundamental processes involved in frequency comb generation have been well tested \cite{udem1999,stenger2002,zimmermann2004,martin2009}, it remains important to test for instability contributions and systematic frequency offsets that may be present in the practical implementation of any particular comb setup, effects which may vary from system to system. Such tests are crucial for a proper 
justification of the uncertainty budgets quoted for absolute optical frequency measurements and optical frequency ratio measurements. If the combs are not to limit such measurements, comb-related systematic uncertainties for optical-optical and optical-microwave comparisons must be below those of the systematic uncertainties of state-of-the-art optical clocks and Cs fountains, currently $6 \times 10^{-18}$ \cite{bloom2014} and $1.1 \times 10^{-16}$ \cite{heavner2014} respectively.  

In this work we compare two combs based on significantly different technologies, with different mode-locking mechanisms and different repetition rates. One system is based on a mode-locked Ti:sapphire laser and the other on a mode-locked Er-doped fibre laser. 
Previous work which (indirectly) compared these two types of comb technology demonstrated agreement at the
2.9$ \times 10^{-19}$ (2$\sigma$) level~\cite{coddington2007}. In other state-of-the-art comb comparisons, two fibre comb systems have displayed agreement at the level of 6$\times 10^{-18}$ \cite{grosche2008} and several different Ti:Sapphire systems have shown agreement at the level of 7$\times 10^{-20}$ (2$\sigma$) \cite{ma2004b,ma2007}. 
In our work we demonstrate agreement between the two comb systems below 3$\times 10^{-21}$ when measuring the ratio of two optical frequencies. To the best of our knowledge this is the highest level of agreement reported between independent fibre and Ti:Sapphire combs and also far surpasses the projected systematic uncertainties of optical clocks. When measuring an optical frequency against a microwave reference the two combs show agreement below 5$\times 10^{-18}$, significantly below the systematic uncertainty of even cryogenic Cs fountains.  
The two combs have recently been used to measure the absolute frequencies of our Sr$^{+}$ and Yb$^{+}$ optical clocks \cite{barwood2014,godun2014} and the optical frequency ratio between two optical clock transitions in Yb$^{+}$~\cite{godun2014}.

The transfer oscillator scheme of Telle \textit{et al.}~\cite{telle2002} is used, in conjunction with synchronous counting, to remove residual comb noise from our measurements. Importantly, several solutions are also implemented to eliminate deviations from 1/$\tau$ white phase noise characteristics on our comb systems, which would otherwise degrade the measurement precision at a given averaging time. For optical frequency ratio measurements between continuous wave (CW) lasers operating at different wavelengths, we ensure that both wavelengths travel together down any unstabilised lengths of fibre. In free space sections, the two wavelengths also follow common paths as far as possible. Frequently observed instability contributions from the erbium doped fibre amplifiers (EDFA) used in fibre based combs \cite{grosche2008,nakajima2010,hagemann2013} are avoided by using a common photonic crystal fibre (PCF) and amplifier branch to observe beats with the two lasers. Although these solutions are implemented for the specific case of an optical frequency ratio measurement between 934 nm and 871 nm lasers used to probe the two optical clock transitions in $^{171}$Yb$^{+}$, it is envisaged that such schemes will also be applicable to other optical frequency ratios of interest when comparing the optical clocks under development at NPL.

The paper is organised as follows. In section~2, the comb systems are described, including their stabilisation schemes and the counting schemes used to measure optical frequencies against a microwave reference and the ratio of two optical frequencies. In section~3, comparisons between optical frequency measurements made using the two comb systems when referenced to a common microwave reference are presented. The limitations of the systems when operating in this manner are discussed, including absolute traceability to the SI second. In section~4 the focus moves to comparisons between optical frequency ratio measurements from the two comb systems, and the limitations of the systems when operating in this manner are also discussed. Section~5 contains conclusions and some perspectives for future development of the comb systems.

\section{Femtosecond frequency comb systems}

\begin{figure}
\begin{center}
\includegraphics[width=8.2cm]{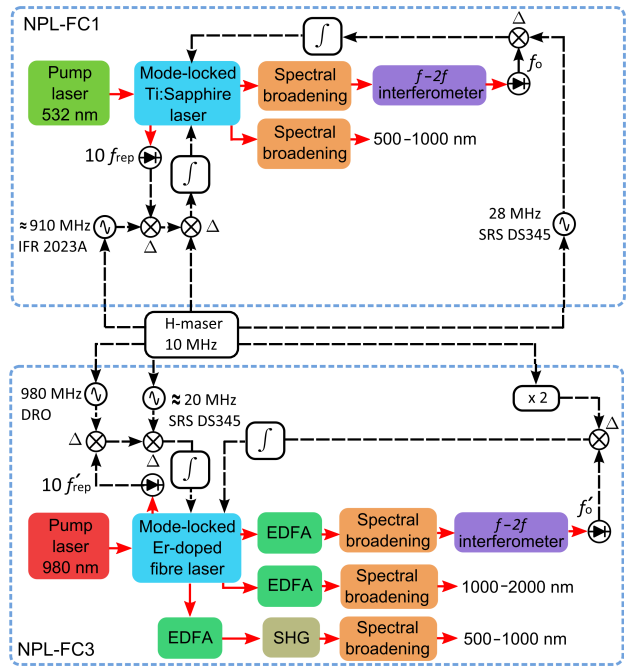}
\caption{The optical branches and stabilisation schemes for the two femtosecond comb systems NPL-FC1 and NPL-FC3. Solid lines indicate optical links and dotted lines indicate electronic links. Filters and amplifiers are omitted from the stabilisation schemes for clarity. When locked, the repetition rates and CEO frequencies of the two combs are $f_{\rm{rep}} \approx$  90~MHz, $f_{\rm{rep}} ^{\prime} \approx$ 100~MHz, $f_{0}$ = 28~MHz and $f_{0}^{\prime}$ = 20~MHz.}
\label{combs}
\end{center}
\end{figure}

The two femtosecond combs used in this work, hereafter referred to as NPL-FC1 and NPL-FC3, are co-located in the same temperature-controlled ($\pm$100~mK) laboratory. Throughout this paper we follow the convention that primed and un-primed terms refer to quantities from NPL-FC3 and NPL-FC1 respectively. Schematics of the optical branches and stabilization schemes of the two comb systems are shown in figure~\ref{combs}. 

\subsection{NPL-FC1}

NPL-FC1 is based on a Kerr-lens mode-locked Ti:Sapphire laser, which was built from a kit supplied by KMLabs and which uses a prism pair for intra-cavity dispersion compensation. The Ti:sapphire laser is pumped with 5 W of 532 nm light from a Coherent Verdi V5 laser and generates a pulse train with an average output power of 650 mW and a repetition rate $f_{\rm{rep}}$ of around 90 MHz. Typically the comb spectrum is centred at 820 nm with a full width at half maximum (FWHM) of 30~nm. 

A fraction of the light from the Ti:Sapphire laser is directed to two independent PCFs for broadening of the comb spectrum. The power and polarisation of the light entering the first PCF are adjusted to generate a spectrum which is peaked at 532 nm and 1064 nm for subsequent detection of the carrier-envelope offset (CEO) frequency $f_{0}$ using the $f-2f$ scheme \cite{jones2000}. Fundamental comb lines at 1064 nm are frequency-doubled to 532 nm in a periodically poled KTiOPO$_{4}$ crystal and heterodyned against fundamental comb lines at 532 nm on an avalanche photodiode (APD). The $f-2f$ interferometer uses a common-path optical arrangement, exploiting a single double-passed Wollaston-prism to compensate for the difference in group delay between the 1064~nm and 532~nm light~\cite{tsatourian2010}. This scheme gives reduced interferometer-induced phase noise on the CEO beat signal compared to more commonly used $f-2f$ arrangements. 

The spectrum from the second PCF is adjusted to optimise the signal-to-noise ratio (SNR) of the beat note(s) $f_{\rm{beat}}$ between the CW laser(s) and the closest comb line(s). The beat detection system employs a diffraction grating to spatially disperse the comb light, minimising the number of unwanted comb lines incident on the subsequent APD(s). Typically a beat SNR of at least 30 dB in a resolution bandwidth (RBW) of 250 kHz is achievable across the range from 500 nm to 1000 nm.
The Ti:Sapphire laser and all free-space optics used for the spectral broadening, $f-2f$ interferometer and beat detection setup are enclosed in separate boxes to reduce the effects of air currents and acoustic noise. 

\subsection{NPL-FC3}

NPL-FC3 is a commercially available (Menlo Systems FC1500) optical frequency comb with a repetition rate $f_{\rm{rep}}^{\prime}$ of around 100 MHz. This comb has been mounted on a wheeled aluminium frame so that it is readily transportable and has previously been described in~\cite{walton2008}. The system is based around an Er-doped fibre laser which is mode-locked via the nonlinear-polarisation-rotation mechanism. The fundamental comb spectrum is centred at 1550 nm and spans around 80 nm FWHM. This light is split between three branches, each amplified with an independent EDFA. 

The comb light in the first branch is broadened to span an octave using a highly nonlinear fibre and is subsequently used to detect the CEO frequency $f_{0}^{\prime}$ using the Menlo Systems $f-2f$ common-path interferometer scheme \cite{menlopatent}. In the second branch, the comb light is broadened in highly nonlinear fibre to supply a spectrum spanning the range 1--2~$\mu$m for frequency measurements in the infra red (IR). In the third branch, the comb light is launched into free space, frequency doubled to 775 nm in a periodically poled lithium niobate crystal and finally coupled through a PCF, supplying a spectrum spanning the range 500--1000 nm for frequency measurements in the near infra red (NIR) to visible. The comb spectra from the different branches can be adjusted independently by squeezing the EDFA fibres by varying degrees.  
In this way it is possible to generate a beat note $f_{\rm{beat}}^{\prime}$ with greater than 30 dB SNR (250 kHz RBW) between the CW laser(s) and closest comb line(s) across the wavelength range 500 nm to 2 $\mu$m. Beat detection is carried out in free space in an identical manner to NPL-FC1. Throughout this work, as the CW lasers are in the NIR, only the 500--1000 nm branch of NPL-FC3 is used. 

\subsection{Stabilisation}

Many previous two-comb comparisons have involved tight phase locking of a single comb mode (and hence all comb modes) to an ultra-stable laser local oscillator (LO) \cite{ma2004b,coddington2007,ma2007}. 
%This ensures that comb measurements can be made without being degraded by the instability of the comb lines. 
In this work the comb modes are only loosely locked to a microwave LO and the transfer oscillator scheme of Telle \textit{et al.}~\cite{telle2002}, in conjunction with synchronous counting, is used to remove the residual instability of the comb modes from the measurements. This exploits the correlations in phase that exist between the radio frequency (RF) beat note of an optical reference with a comb line, and both the comb repetition rate and CEO frequency. Previous studies of comb stability and systematic uncertainty using the transfer oscillator scheme have been presented \cite{stenger2002,grosche2008,benkler2005,nicolodi2014}; however the work reported here significantly improves upon previously reported systematic uncertainties.

The repetition rates and CEO frequencies of NPL-FC1 and NPL-FC3 are stabilised to RF references using phase-lock-loop (PLL) electronics. All reference frequencies are derived from a common 10 MHz signal from a H-maser (Sigma Tau MHM-2010) which minimises the frequency drift of the comb lines. The H-maser resides in a separate laboratory, which is linked to both the combs laboratory and the NPL caesium fountain laboratory with high quality RF cabling (Andrew FSJ1-50A) and low-noise 10 MHz distribution amplifiers (Spectradynamics HPDA-15RM, DS-100 and Time \& Frequency Solutions FDA-1050). Figure~\ref{combs} displays the locking schemes for the two systems. To improve phase sensitivity, both combs use the tenth harmonic of the repetition rate for locking. The frequencies of significance are $f_{\rm{rep}} \approx$  90~MHz, $f_{\rm{rep}} ^{\prime} \approx$ 100~MHz, $f_{0}$ = 28~MHz and $f_{0}^{\prime}$ = 20~MHz. The mode spacings $f_{\rm{rep}}$ and $f_{\rm{rep}}^{\prime}$ can be tuned independently by changing the frequencies of a 910~MHz synthesiser (IFR 2023A) or a 20~MHz synthesiser (SRS DS345) respectively. 

Since the transfer oscillator scheme is implemented on both systems, tight locking of the repetition rates and CEO frequencies is not required and the servos all have moderately low bandwidths of $<$10 kHz on both combs. For NPL-FC1 $f_{0}$ can be tuned and locked using a mirror mounted on a split-piezo actuator in the arm of the laser containing the prism pair. This allows the end mirror of the Ti:Sapphire cavity to be tilted, which modifies the intra cavity dispersion. For NPL-FC3 $f_{0}^{\prime}$ can be tuned and locked by changing the intra cavity dispersion via the pump power. For both combs the repetition rate can be finely tuned and locked by adjusting the cavity length with a mirror mounted on an ordinary piezo actuator, and can be coarsely tuned by translating this mirror mount using a motorised stage. 

\subsection{$f_{\rm{rep}}$ counting}
\label{repcounting}

\begin{figure}
\begin{center}
\includegraphics[width=8.2cm]{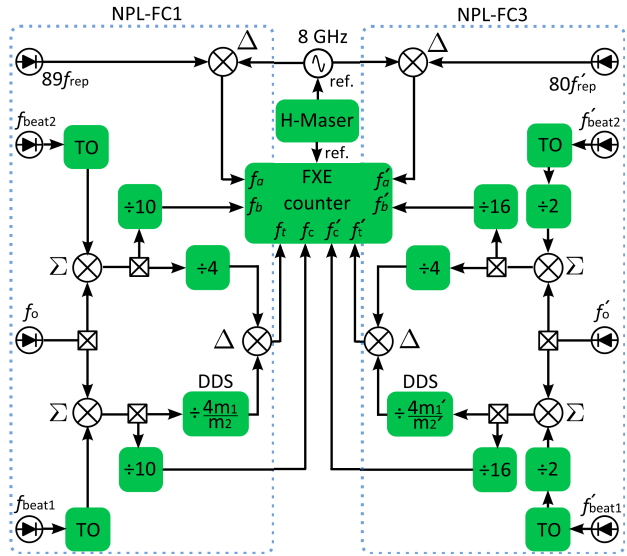}
\caption{The counting schemes for both comb systems. TO: Tracking oscillator, DDS: Direct digital synthesiser. Primed and un-primed terms indicate quantities from NPL-FC1 and NPL-FC3 respectively. Filters and amplifiers are omitted for clarity.}
\label{countingratio}
\end{center}
\end{figure}

To efficiently remove the fluctuations of the loosely stabilised comb modes from optical frequency measurements, it is important that the repetition rate is monitored with high enough resolution such that the counter noise floor is well below these fluctuations. To achieve this, a high harmonic of the repetition rate is detected at $\sim$8 GHz (89$f_{\rm{rep}}$ for NPL-FC1, 80$f^{\prime}_{\rm{rep}}$ for NPL-FC3) using a fast low-noise photodetector (Discovery Semiconductors DSC40S). This $\sim$8~GHz signal is amplified by 40~dB then subsequently mixed down to $\sim$1.5~MHz via a fixed maser-referenced 8~GHz synthesiser (Agilent E8247C) which is common to both combs. The mixed down frequencies $f_{\rm{a}}$ and $f^{\prime}_{\rm{a}}$ from each comb system are counted synchronously on individual channels of a 12-channel counter, referenced to the same H-maser. The counter is made up of three cascaded four-channel $\Pi$-type dead-time-free K+K FXE counter boards \cite{kramer2004}. This counting scheme is illustrated in the upper part of figure \ref{countingratio}.

\subsection{$f_{0}$ and $f_{\rm{beat}}$ counting}
\label{foandfbeat}

To suppress fluctuations of the CEO frequencies from optical frequency measurements the frequencies $f_{0}$ ($f_{0}^{\prime}$) and $f_{\rm{beat}}$ ($f_{\rm{beat}}^{\prime}$) are added on each comb system using double balanced mixers (DBMs), with all beat signs chosen to be positive. This follows the transfer oscillator scheme of Telle \textit{et al.}~\cite{telle2002} and is illustrated in figure \ref{countingratio}. For the case of NPL-FC3, when using the frequency-doubled branch of the comb, fluctuations of $f_{0}^{\prime}$ appear a factor of two larger on the optical beats $f_{\rm{beat}}^{\prime}$ and digital divide-by-two stages are used before mixing to account for this. The two sum frequencies from the two comb systems are divided by either 10 or 16, to bring them into the operable range of the FXE counters, giving frequencies $f_{b}$, $f_{c}$, $f_{b}^{\prime}$ and $f_{c}^{\prime}$. These are then counted on additional channels of the FXE counters, synchronously with the $f_{\rm{a}}$ and $f_{\rm{a}}^{\prime}$ signals. The optical frequencies of the CW lasers 1 and 2 measured by the two combs can then be calculated using the following four equations:
\begin{eqnarray}
f_{1} & =  m_{1}  f_{\rm{rep}} + f_{0} + f_{\rm{beat1}}   \nonumber\\
& =  m_{1}  \left(\frac{f_{a} + 8 \, \rm{GHz}}{89} \right) + 10 f_{c},
\end{eqnarray}
\begin{eqnarray}
f_{2} & =  m_{2}  f_{\rm{rep}} + f_{0} + f_{\rm{beat2}}\nonumber\\
& =  m_{2}  \left(\frac{f_{a} + 8 \, \rm{GHz}}{89} \right) + 10 f_{b},
\end{eqnarray}
\begin{eqnarray}
f_{1}^{\prime} & =  m_{1}^{\prime}  f_{\rm{rep}}^{\prime} + 2f_{0}^{\prime} + f_{\rm{beat1}}^{\prime}\nonumber\\
& =  m_{1}^{\prime}  \left(\frac{f_{a}^{\prime} + 8 \, \rm{GHz}}{80} \right) + 32 f_{c}^{\prime},
\end{eqnarray}
\begin{eqnarray}
f_{2}^{\prime} & =  m_{2}^{\prime}  f_{\rm{rep}}^{\prime} + 2f_{0}^{\prime} + f_{\rm{beat2}}^{\prime}\nonumber\\
& =  m_{2}^{\prime}  \left(\frac{f_{a}^{\prime} + 8 \, \rm{GHz}}{80} \right) + 32 f_{b}^{\prime},
\end{eqnarray}
where $m_{1}$, $m_{1}^{\prime}$, $m_{2}$ and $m_{2}^{\prime}$ are the numbers of the comb modes used to generate beat notes $f_{\rm{beat1}}$, $f_{\rm{beat1}}^{\prime}$, $f_{\rm{beat2}}$ and $f_{\rm{beat2}}^{\prime}$ respectively. By synchronously counting all frequencies from the two combs the residual fluctuations of the repetition rates of the two combs are also removed from the measurements. 

To remove broadband noise and ensure accurate counting the beats are pre-filtered and amplified using tracking oscillators (TO) with bandwidths of around 500 kHz. The CEO frequencies typically have a 40 dB or greater SNR (250 kHz RBW) on both combs and do not require such pre-filtering.

\subsection{Generating a transfer beat $f_{t}$}
\label{transfer beat}

Figure \ref{countingratio} also shows the counting scheme for measuring the ratio of two spectrally separated CW laser frequencies on either system. This is an extension of the transfer oscillator scheme described in section \ref{foandfbeat} and allows transfer beats $f_{t}$ and $f_{t}^{\prime}$ to be generated between the two CW lasers that are free from the residual fluctuations of the repetition rates and CEO frequencies of the two combs.
For example, for the case of NPL-FC1, to suppress the fluctuations of $f_{\rm{rep}}$ a direct digital synthesiser (DDS) is used to scale the fluctuations of $f_{\rm{beat1}}+f_{0}$ by the ratio of the comb mode numbers $m_{1}/m_{2}$ that the two CW lasers are beating against. A common division by four is also introduced on both arms to bring the division ratio into the range that can be produced by the DDS. The difference frequency between these two arms, or transfer beat $f_{t}$, is then formed on a final DBM:
\begin{equation}
f_{t} = \frac{1}{4} \frac{m_{2}}{m_{1}} \left( f_{\rm{beat1}} + f_{0}  \right) - \frac{1}{4} \left( f_{\rm{beat2}} + f_{0}   \right) ,
\end{equation}
and contains only fluctuations of the two CW laser frequencies $f_{1}$ and $f_{2}$ since
\begin{eqnarray}
f_{t}  & =  \frac{1}{4} \left[ \frac{m_{2}}{m_{1}} \left( f_{1} - m_{1}f_{\rm{rep}}  \right) - \left( f_{2} - m_{2}f_{\rm{rep}}   \right) \right] \nonumber\\
& =  \frac{1}{4} \left( \frac{m_{2}}{m_{1}} f_{1} - f_{2} \right)  .
\end{eqnarray}

The transfer beats $f_{t}$ and $f_{t}^{\prime}$ from the two combs  are counted synchronously on independent channels of the FXE counters. The optical frequency ratios measured using NPL-FC1 and NPL-FC3 can be determined from the transfer beats via the respective equations
\begin{equation}
\frac{f_{1}}{f_{2}} =  \frac{m_{1}}{m_{2}} \left( 1+\frac{4 f_{t}}{f_{2}} \right)\label{ratio1}
\end{equation}
and
\begin{equation}
\frac{f_{1}^{\prime}}{f_{2^{\prime}}}  =  \frac{m_{1}^{\prime}}{m_{2}^{\prime}} \left( 1+\frac{8 f_{t}^{\prime}}{f_{2}^{\prime}} \right).\label{ratio2}
\end{equation}
The second terms in the brackets of equations \ref{ratio1} and \ref{ratio2} are below $ 10^{-7}$ and as a result $f_{t}$ and $f_{2}$ (or $f_{t}^{\prime}$ and $f_{2}^{\prime}$) need only be counted/known to fractional uncertainties of $10^{-11}$ to be able to measure optical frequency ratios with uncertainties below $10^{-18}$, suitable for most present-day optical clock comparisons.  

The scheme illustrated in figure \ref{countingratio} allows synchronous optical frequency measurements of the two CW lasers to be made as well as direct measurements of the optical frequency ratio between them. This gives the option of suppressing the fluctuations of the comb repetition rate from the optical frequency ratio measurements in software post processing, by relying on the synchronous counting of the signals $f_{a}$, $f_{b}$, $f_{c}$ and $f_{a}^{\prime}$, $f_{b}^{\prime}$, $f_{c}^{\prime}$. However, carrying out the suppression in hardware has the added advantage that a transfer beat, containing information about the phase fluctuations of the two CW lasers, can be monitored in real-time on a spectrum analyser.

\section{Optical-to-microwave comparison}

Comparisons were made between NPL-FC1 and NPL-FC3 by synchronously measuring the optical frequency of a common ultra-stable clock laser at 934 nm, stabilised using a high-finesse ULE cavity, against a common microwave reference, using the two comb systems. The 934 nm light was sent between the Yb$^{+}$ optical clock laboratory and the combs laboratory via a 50 m polarisation-maintaining (PM) phase-noise-cancelled fibre link. Once in the combs laboratory the 934 nm light was split between the two combs using a single-mode PM 50:50 fibre beam splitter. The FXE counters were used to count all frequencies from both combs synchronously with a 1 s gate time.

\subsection{Residual instability}

\begin{figure}
\begin{center}
\includegraphics[width=8cm]{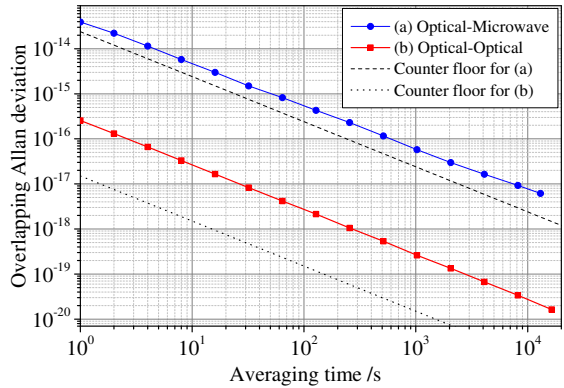}
\caption{Overlapping Allan deviation of (a) a continuous 52\,000 s data set showing the fractional frequency instability between the two comb systems when measuring the same optical frequency against the same maser-referenced microwave oscillator; (b) a continuous 72\,000 s data set showing the residual stability of the difference between two independent combs measuring the same optical frequency ratio. The counter noise floors are also displayed.}
\label{comparisonallan}
\end{center}
\end{figure}

Data set (a) in figure \ref{comparisonallan} shows the overlapping Allan deviation of the fractional difference between the measured optical frequencies from the two comb systems over a continuous data collection period of approximately 52\,000 s.
%Rem2_110706_Frequ_04 & 110707_Frequ_01.txt 
The instability follows a $4 \times 10^{-14} \: \tau^{-1}$ line and
continues down to $6\times10^{-18}$ after 13\,000 s of averaging. 
In order to achieve this performance it was critical to identify the non-common thermal expansions and mechanical stresses of the two output ports of the 50:50 fibre beam splitter (both of which are 1 m in length and not phase stabilised) as well as the non-common thermal effects between the two combs themselves. Such effects are expected to limit the long-term agreement between the two combs when making optical frequency measurements. Although both comb systems are situated in the same temperature-controlled laboratory, the combs and most of the two lengths of fibre of the 50:50 splitter outputs reside in independent boxed enclosures, which can undergo non-common temperature drifts due to heat-generating equipment. Using temperature probes we made sure that these enclosures had both reached a stable internal temperature before taking measurements, otherwise floors in the Allan deviation and frequency offsets were observed between the combs at the low parts in 10$^{17}$ level. Using fibre properties from \cite{tateda1980}, this is consistent with the expected fractional frequency change in 1 m of fibre whose temperature changes by 1 K over an hour. It was found that mechanical vibrations of the fibres from passing air currents are also a critical problem. The fibre lengths are secured and shielded using acoustic foam or placed inside enclosures, otherwise short-lived but large offsets between the combs at the $10^{-15}$ level could be observed. These effects were investigated in a separate experimental study by monitoring a heterodyne beat between the two outputs ports of the 50:50 fibre splitter.

The thermal expansion of the microwave cables and components used to carry the 8 GHz signals is another potential source of frequency instability over long time scales. Non-common temperature fluctuations of the cables, due to different routings, could lead to frequency fluctuations between the combs.
%at the low 10$^{-17}$ level 
We carried out an experimental study and found no such effects larger than the $10^{-18}$ level from the two 8 GHz detection schemes.

In the two-comb comparisons the fluctuations of the H-maser are common and the residual instability between the two systems is dominated by the $f_{\rm{rep}}$ ($f_{\rm{rep}}^{\prime}$) detection schemes. Phase noise measurements indicate that the white phase noise floor introduced by the three 8 GHz amplifiers used to increase the level of the 89$f_{\rm{rep}}$ and 80$f_{\rm{rep}}^{\prime}$ signals is $-$130 dBc/Hz. This corresponds to a frequency instability of $\sim$2 $\times 10 ^{-14}$ at 1 s for each comb system in a bandwidth of $\sim$1~MHz, in good agreement with the observed instability for the two-comb comparison. The combined residual frequency instability of the two 8~GHz repetition rate detection schemes was also directly measured in a separate study to be 4$\times 10^{-14} \: \tau^{-1}$, confirming that this limits the short term stability for optical frequency measurements relative to a microwave reference. The counter resolution limit for the $f_{\rm{rep}}$ ($f_{\rm{rep}}^{\prime}$) detection (without any slew rate enhancement), measured using a dummy 1.5~MHz signal from a synthesiser with equivalent signal level, is also shown in figure \ref{comparisonallan} for reference.

Over all timescales, the residual instability of optical-microwave comparisons achievable from the two combs lies significantly below the instability of the H-maser reference, which typically follows a frequency stability of 5 $\times 10^{-13} \: \tau^{-1}$ down to a floor at the $10^{-15}$ level. 
This residual instability also lies below the instability of the NPL Cs fountain NPL-CsF2 \cite{szymaniec2010} over all time scales. The combs themselves do not therefore present a stability limitation for absolute optical frequency measurements.     

\subsection{Cycle slip detection}

If the SNRs of the $f_{\rm{beat}}$ or $f_{\rm{beat}}^{\prime}$ signals fall below about 30 dB (250 kHz RBW), the PLLs of the TOs can be susceptible to cycle slips. To detect potential slips, two independent TOs with slightly different bandwidths are locked to the same beat signal on each comb system. This results in the two TOs being affected differently by the noise on the beat, seeing different amounts of integrated power. The two TO frequencies are then counted synchronously during optical frequency measurements. As the frequency difference between the two TOs has a residual instability much less than 1 Hz at 1 s, cycle slips are easily detected by looking for differences between the two TO frequencies corresponding to an integer number of hertz in a 1 s gate time. Such data points can then be removed before analysis. 

\subsection{Systematic uncertainty}
\label{accuracy absolute}

\begin{figure}
\begin{center}
\includegraphics[width=8cm]{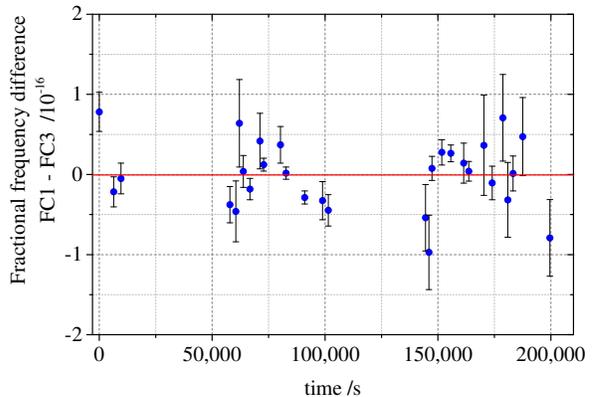}
\caption{Comparisons of NPL-FC1 and NPL-FC3 optical frequency measurements relative to a common microwave reference, spanning 4 days.}
\label{comparisonfreqlong}
\end{center}
\end{figure}

To understand long-term behaviour several data sets taken over a period of 4 days were analysed, and the fractional frequency differences between the two combs are plotted in figure \ref{comparisonfreqlong}. All data sets were continuous with duration longer than 1000 s, containing no cycle slips. Error bars are based on the extrapolated Allan deviation for each data set. The weighted mean of these data is $-$0.5  $\times 10^{-18}$ with a relative standard uncertainty of 4.8$\times 10^{-18}$. This demonstrates
that there is no detectable frequency offset between the two combs at the 5$\times10^{-18}$ level, which is limited by the measurement statistics. 

As the two combs share the same H-maser-referenced 8 GHz synthesiser for repetition rate counting (see figure \ref{countingratio}) the comb comparison results are insensitive to possible frequency errors introduced by the synthesiser in multiplying the original 10 MHz H-maser signal to 8 GHz. Such frequency offsets may arise due to slow temperature drifts within the synthesiser circuitry. On both comb systems these errors will typically dominate the overall uncertainty of an absolute frequency measurement, due to the large multiplication factor of the repetition rate in scaling to the optical frequency. Potential errors in the 8 GHz synthesis are detected by taking part of the 8 GHz output signal from the synthesiser and dividing it by 800 using a chain of low-noise digital frequency dividers ( $\div$2 $\div$2  Hittite HMC361, $\div$4 Hittite HMC-394LP, $\div$50 Zarlink Semiconductor SP8400). During all absolute optical frequency measurements the phase difference between this 10 MHz signal generated by division and the original 10 MHz H-maser signal is logged on a phase comparator (TimeTech TAF-0502) at 10 s intervals. The phase comparison is carried out in the H-maser laboratory, as opposed to the combs laboratory. This ensures that the phase comparison analysis includes any errors imparted to the 10 MHz signal by the RF link and distribution amplifiers between the two laboratories. It thus gives a conservative measure of potential frequency offsets, containing not only errors introduced by the 8 GHz synthesiser but also potential errors from the digital division and the return path of the RF link, which are not actually involved in a frequency measurement.
%Some sample phase comparison data is shown in Figure \ref{10MHz} along with the associated Allan deviation. 
It is found that uncertainties associated with the 8 GHz synthesis and 10 MHz distribution are typically below $1 \times 10^{-16}$ for the long measurement campaigns required when measuring absolute frequencies, limited by the noise floor of the phase comparator. This is below the systematic frequency uncertainty of NPL-CsF2 (see section \ref{traceability}) and therefore does not pose a significant limitation to absolute frequency measurements at present.

\subsection{Traceability to the SI second}
\label{traceability}

To achieve traceability to the SI second, during optical frequency measurements with either comb system the frequency offset of the H-maser from 10\,MHz is determined by reference to the NPL Cs fountain NPL-CsF2, which is described in \cite{szymaniec2010}.  
This frequency offset is averaged and reported every 1600 s. The statistics of NPL-CsF2 average down as $3 \times 10^{-13} \tau^{-1/2}$ when it is operated in alternating density mode \cite{szymaniec2010}, which determines the duration of data taking required to reach a given statistical uncertainty.  

With exact temporal overlap between the fountain and comb data-taking periods, the H-maser acts as a perfect transfer oscillator and its frequency fluctuations cancel out in the absolute optical frequency measurements. However, in practice there will be short periods of dead time and offsets between the measurement intervals. The effect of these can be accounted for by inflating the statistical error bars using mathematical expressions that take account of the noise behaviour of the H-maser \cite{yu2007}.  
When these periods are short compared to the total duration of overlapping data, these modifications are negligible compared to the statistical uncertainty associated with the fountain. Periods of dead time are typically associated with the optical clocks; the duty cycle of NPL-CsF2 is approximately 90$\%$.

The systematic uncertainty of NPL-CsF2 has recently been evaluated as $2.0\times 10^{-16}$ \cite{szymaniec2014}. Taking out the uncertainty associated with correcting to the geoid, this becomes 1.9 $\times 10^{-16}$ and sets a lower limit to the uncertainty of absolute optical frequency measurements at NPL. One must also consider the gravitational red shift associated with the difference in height between the optical standard and the time-averaged height of the Cs atoms in NPL-CsF2. For all the NPL optical clocks, this difference in height is less than 1 m and can easily be determined with an uncertainty at the 10 cm level. This corresponds to a required correction below $1 \times 10^{-16}$ with an uncertainty at the $1 \times 10^{-17}$ level.

\section{Optical-to-optical comparison}

Comparisons were also made between NPL-FC1 and NPL-FC3 by measuring the same optical frequency ratio, between the two ultra-stable clock lasers at 934 nm and 871 nm, synchronously using the two comb systems.
The 934 nm and 871 nm light was sent between the Yb$^{+}$ optical clock laboratory and the combs laboratory via independent 50~m PM phase-noise-cancelled fibre links. 
Once in the combs laboratory the 934 nm and 871 nm lasers were launched into free space, combined using a dichroic mirror, and coupled into the same input port of a single mode 50:50 PM fibre splitter. The two output ports of the splitter thus divert light at both wavelengths to NPL-FC1 and NPL-FC3, where they are combined with comb light using free space optics. In this way the two wavelengths follow a common path on each comb up to the diffraction grating, where they are diffracted at different angles onto independent APDs for beat detection at 934 nm and 871 nm. The FXE counters were used to count the transfer beats $f_{t}$ and $f_{t}^{\prime}$ from both combs synchronously with a 1 s gate time.

\subsection{Residual instability}

Data set (b) in figure \ref{comparisonallan} shows the overlapping Allan deviation of the difference between the measured optical frequency ratios from the two comb systems measured over a continuous 72\,000~s period. 
The instability follows a $2.6\times 10^{-16} \: \tau^{-1}$ line reaching $1.6 \times 10^{-20}$ after 16\,000~s with no indications of a floor being reached within the accessible time scales of the experiment. In other works involving multi-branch fibre combs, instability plateaus at the $10^{-15}$ level have been observed out to about 10~s before averaging down at longer time scales. These plateaus are due to differential fibre-related phase noise in EDFAs~\cite{grosche2008,nakajima2010}. In our experiments, the use of a single EDFA to generate comb light at both 934~nm and 871~nm suppresses the effect of fibre-induced phase fluctuations which, to lowest order, scale linearly with optical frequency. In the same way, coupling both the 934~nm and 871~nm lasers into the same fibre before splitting the light to the two combs means the ratio between the two laser frequencies is not affected by changes in the unstabilised lengths of fibre that make up the splitter. The fibre-induced relative frequency noise between the two ports of the fibre beam splitter was measured as $1 \times 10^{-18}$ at 1000~s in a separate study with a single CW laser, but for the ratio measurement these fluctuations are efficiently suppressed below the $2.6 \times 10^{-16} \: \tau^{-1}$ line in figure \ref{comparisonallan}. 

In other recent work the CEO frequency has also been detected using the same EDFA branch as that used to generate the beats with the optical references~\cite{nicolodi2014}, but in this work we find that this is unnecessary. Again this is to be expected for fibre-induced phase fluctuations which scale linearly with optical frequency, affecting $2(f_{0}+nf_{\rm rep})$ and $f_{0}+2nf_{\rm rep}$ almost equally and making the $f_{0}$ beat signal insensitive to these fluctuations to a high degree. Although one does expect ultimately to see contributions to the comb instabilities due to phase fluctuations which scale with order higher than linear with optical frequency, such effects are at a level below that relevant for optical clock comparisons. 

The typical SNR of 30~dB (250~kHz RBW) of each of the four beat notes $f_{\rm{beat1}}$, $f_{\rm{beat2}}$, $f^{\prime}_{\rm{beat1}}$ and $f^{\prime}_{\rm{beat2}}$ corresponds to a white phase noise floor of approximately $-$80~dBc/Hz which, for the TO bandwidths used ($\sim$500~kHz), results in a residual frequency instability of about $1 \times 10 ^{-16} \: \tau^{-1} $ for each TO branch. This was confirmed experimentally, using dummy signals with white noise. Adding the contributions from the four TOs in quadrature, one expects a residual instability of the ratio difference of about $2 \times 10 ^{-16}$ at 1 s, in reasonable agreement with that observed experimentally. Figure \ref{comparisonallan} also shows the expected noise floor of the counters for reference. 

As discussed in section \ref{transfer beat} it is also possible to suppress the fluctuations of $f_{\rm{rep}}$ on the two combs in software post processing, in order to determine the optical frequency ratio. Such an analysis was carried out and shows comparable performance for the ratio difference instability data, in this case following a line of $2.8 \times 10^{-16} \: \tau ^{-1}$.

The residual instability of optical-optical comparisons achievable from the two combs over all time scales lies significantly below the present instabilities of the optical clocks under development at NPL, therefore the combs do not currently limit the stability achievable when comparing the different optical clocks.  
 
\subsection{Cycle slip detection}

\begin{figure}
\begin{center}
\includegraphics[width=8cm]{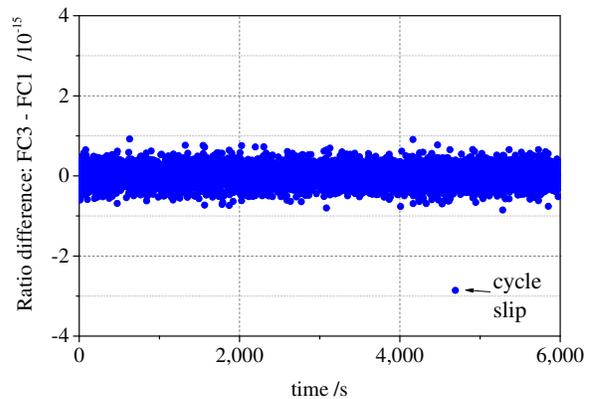}
\caption{A short section of data showing the difference in measured optical frequency ratio between the two comb systems. The presence of a single cycle slip on one of the four TOs is clearly visible.}
\label{comparisonfreq}
\end{center}
\end{figure}

For cycle slip detection, rather than using two TOs on each of the four beats $f_{\rm{beat1}}$, $f_{\rm{beat2}}$, $f^{\prime}_{\rm{beat1}}$ and $f^{\prime}_{\rm{beat2}}$, the short term instability of the residual difference between both combs for optical-optical comparisons is low enough that cycle slips can be detected directly. Equations \ref{ratio1} and \ref{ratio2} indicate that a slip of 1 Hz on any of the four TOs corresponds to a difference in the frequency ratio measured by  the two combs of approximately 3$\times$10$^{-15}$ in a 1 s gate time. This is illustrated in figure \ref{comparisonfreq} which shows a single cycle slip appearing in a short set of comb comparison data. Using this method, data points which contain cycle slips are removed before analysis.

\subsection{Systematic uncertainty}

\begin{figure}
\begin{center}
\includegraphics[width=8cm]{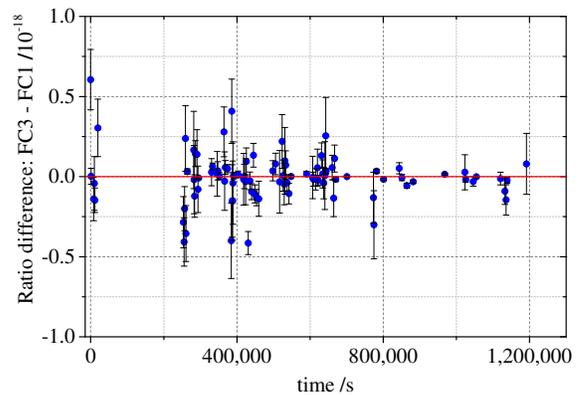}
\caption{Comparisons of NPL-FC1 and NPL-FC3 optical frequency ratio measurements spanning around 14 days. The red line represents the weighted mean of the data points.}
\label{comparisonratiolong}
\end{center}
\end{figure}

Figure \ref{comparisonratiolong} shows the offset between synchronous optical frequency ratio measurements made using the two comb systems for data sets collected over two weeks. All data sets were continuous with duration longer than 1000 s and containing no cycle slips. Error bars are based on the extrapolated Allan deviation for each data set. The weighted mean of these data is 
$-$0.4$\times 10^{-21}$ with a relative standard uncertainty of 2.7$\times 10^{-21}$. Since the ratio measured in this case is approximately 0.93, this indicates that there is no detectable systematic offset between the two comb systems at a fractional level of $2.9\times 10^{-21}$, with this limit set by the measurement statistics. 

The division by $4 m_{1} / m_{2}$ and $4 m_{1}^{\prime} / m_{2}^{\prime}$ on both comb setups can only be realised to within a certain accuracy due to the finite tuning word length of the DDSs used (Analogue Devices AD9912). These support a 48-bit frequency tuning word, resulting in a known systematic error at the $10^{-21}$ level for a frequency ratio measurement. The division ratio is different for the two combs, resulting in different DDS errors, which lead to an expected offset of around 2$\times 10^{-21}$ between the two comb systems, just below the statistical uncertainty for the data shown in figure~\ref{comparisonratiolong}.

In this study, the CW lasers used were not always locked to an atomic reference. Instead of using a fixed value for $f_{2}$ and $f_{2}^{\prime}$ in equations \ref{ratio1} and \ref{ratio2} respectively, this frequency was synchronously counted on the two combs along with the transfer beat frequencies during all measurements. Because frequencies $f_{t}$ ($f_{t}^{\prime}$) and $f_{2}$ ($f_{2}^{\prime}$) are compared against a common H-maser, which is used as the reference for the FXE counters, optical frequency ratio measurements from either comb become insensitive to the absolute frequency of the H-maser.

\section{Summary and conclusions}

Through two-comb comparisons it has been demonstrated that two independent comb systems NPL-FC1 and NPL-FC3 introduce relative systematic errors below 5$\times10^{-18}$ when measuring optical frequencies relative to a microwave reference and below 3$\times10^{-21}$ when measuring the ratio of two optical frequencies. Even when including potential systematic shifts in the traceability between the combs and the NPL Cs fountain primary standard NPL-CsF2, the uncertainty of absolute frequency measurements remains below the systematic uncertainty of NPL-CsF2. 

The two combs have recently been used to measure the absolute frequencies of the NPL Sr$^{+}$ and Yb$^{+}$ optical clocks \cite{barwood2014,godun2014}. The uncertainty demonstrated for optical frequency ratio measurements is far beyond even the projected systematic uncertainties of optical clocks; the two combs have also recently been used to measure the optical frequency ratio between two clock transitions in Yb$^{+}$~\cite{godun2014}, with an uncertainty dominated by the uncertainty in the systematic shift of the electric quadrupole transition. 

Although the two femtosecond comb systems are situated in the same laboratory, they are based on different laser technologies with different repetition rates and so potential systematic frequency offsets that could occur from spectral broadening or (in the case of NPL-FC3) optical amplification, thermal and acoustic effects are all expected to be uncorrelated between the two systems. Care has been taken to eliminate possible common-mode systematics from air currents and temperature that are related to the CW laser delivery and beat detection; similar fibre routes are avoided and the beat detection sections of the two combs are orthogonal to one another.  

There is currently ongoing development around the globe of next-generation laser LOs with instabilities below $1 \times 10^{-16}$ at short time scales \cite{jiang2011,kessler2012,cole2013,leibrandt2013,bohnet2012}, which promise associated improvements in the short-term instability of optical clocks. However, such oscillators must sometimes operate in specific spectral regions, meaning that a frequency comb is required to transfer the stability of the oscillator to the desired optical clock transition frequency. For the two comb systems investigated here it will therefore become important to improve the residual short term instability achievable for optical-optical comparisons below the current level, such that the spectral fidelity of NPL's next-generation LOs can be preserved during transfer. It is envisaged that improvements to the short-term instability can be obtained for these comb systems by increasing the SNR of the beat notes between the optical references and comb lines. This could be achieved by implementing better mode matching between comb and CW laser light, narrower band filtering of comb lines prior to beat detection or increasing the optical power of  the comb, for example by the injection locking of slave lasers to the comb lines of interest \cite{wu2013}. Alternatively, by stabilising the comb to an optical reference, narrower band filtering of the beat frequencies can be used with reduced TO bandwidths, effectively increasing the SNR of the filtered beat notes \cite{nicolodi2014}.

\ack

The authors would like to thank the NPL Yb$^{+}$ ion trappers for the use of their clock lasers and Giuseppe Marra for useful help and discussions regarding RF electronics and tracking oscillator design. This work was funded by the UK National Measurement System as part of the Electromagnetics and Time Programme.
\section*{References}
\bibliographystyle{unsrt}
\bibliography{CombComparison}

\end{document}